\documentclass[12pt]{article}
\usepackage{epsfig}

\newcommand\pubnumber{RADCOR-2000-58}
\newcommand\pubdate{\today}
\newcommand\hepnumber{hep-ex/0101042}

\def\support{\footnote{This work was supported in part by the US Dept.~of Energy, The
U.S.~National Science Foundation, the
German Bundesminister f\"ur Bildung und Forschung, the Russian Ministry of
Science, and the US-Japan Agreement in High Energy Physics. A.~Steinmetz
acknowledges support by the Alexander von Humboldt Foundation.}}

\def\Title#1{\begin{center} {\Large\bf #1 } \end{center}}
\def\Author#1{\setlength{\baselineskip}{10pt} \begin{center}{\footnotesize #1} \end{center}}
\def\Address#1{\begin{center}{ \it \scriptsize #1} \end{center}}

\newcommand\pubblock{\rightline{\begin{tabular}{l} \pubnumber\\
         \pubdate\\ \hepnumber \end{tabular}}}
\newenvironment{Abstract}{\begin{quotation}  }{\end{quotation}}
\newenvironment{Presented}{\begin{quotation} \begin{center} 
             Presented at the\end{center}
      \begin{center}\begin{large}}{\end{large}\end{center} \end{quotation}}

\makeatletter
\def\section{\@startsection{section}{0}{\z@}{5.5ex plus .5ex minus
 1.5ex}{2.3ex plus .2ex}{\large\bf}}
\def\subsection{\@startsection{subsection}{1}{\z@}{3.5ex plus .5ex minus
 1.5ex}{1.3ex plus .2ex}{\normalsize\bf}}
\def\subsubsection{\@startsection{subsubsection}{2}{\z@}{-3.5ex plus
-1ex minus  -.2ex}{2.3ex plus .2ex}{\normalsize\sl}}

\renewcommand{\@makecaption}[2]{%
   \vskip 10pt
   \setbox\@tempboxa\hbox{\small #1: #2}
   \ifdim \wd\@tempboxa >\hsize     
       \small #1: #2\par          
     \else                        
       \hbox to\hsize{\hfil\box\@tempboxa\hfil}
   \fi}

 \def\citenum#1{{\def\@cite##1##2{##1}\cite{#1}}}
 
\newcount\@tempcntc
\def\@citex[#1]#2{\if@filesw\immediate\write\@auxout{\string\citation{#2}}\fi
  \@tempcnta\z@\@tempcntb\m@ne\def\@citea{}\@cite{\@for\@citeb:=#2\do
    {\@ifundefined
       {b@\@citeb}{\@citeo\@tempcntb\m@ne\@citea\def\@citea{,}{\bf ?}\@warning
       {Citation `\@citeb' on page \thepage \space undefined}}%
    {\setbox\z@\hbox{\global\@tempcntc0\csname b@\@citeb\endcsname\relax}%
     \ifnum\@tempcntc=\z@ \@citeo\@tempcntb\m@ne
       \@citea\def\@citea{,}\hbox{\csname b@\@citeb\endcsname}%
     \else
      \advance\@tempcntb\@ne
      \ifnum\@tempcntb=\@tempcntc
      \else\advance\@tempcntb\m@ne\@citeo
      \@tempcnta\@tempcntc\@tempcntb\@tempcntc\fi\fi}}\@citeo}{#1}}
\def\@citeo{\ifnum\@tempcnta>\@tempcntb\else\@citea\def\@citea{,}%
  \ifnum\@tempcnta=\@tempcntb\the\@tempcnta\else
  {\advance\@tempcnta\@ne\ifnum\@tempcnta=\@tempcntb \else\def\@citea{--}\fi
    \advance\@tempcnta\m@ne\the\@tempcnta\@citea\the\@tempcntb}\fi\fi}
\makeatother

\def\beq{\begin{equation}}
\def\eeq#1{\label{#1}\end{equation}}
\def\eeqn{\end{equation}}

\newenvironment{Eqnarray}%
   {\arraycolsep 0.14em\begin{eqnarray}}{\end{eqnarray}}
\def\beqa{\begin{Eqnarray}}
\def\eeqa#1{\label{#1}\end{Eqnarray}}
\def\eeqan{\end{Eqnarray}}

\let\bar=\overbar

\def\Dslash{\not{\hbox{\kern-4pt $D$}}}
\def\dslash{\not{\hbox{\kern-2pt $\del$}}}

\def\ee{e^+e^-}

\def\msb{{\bar{\ssstyle M \kern -1pt S}}}

\def\lsim{\mathrel{\raise.3ex\hbox{$<$\kern-.75em\lower1ex\hbox{$\sim$}}}}
\def\gsim{\mathrel{\raise.3ex\hbox{$>$\kern-.75em\lower1ex\hbox{$\sim$}}}}

\def\be{\begin{equation}}
\def\ee{\end{equation}}
\def\bea{\begin{eqnarray}}
\def\eea{\end{eqnarray}}
\def\g2{$(g-2)$}
\def\amu{$a_{\mu}$\hskip0.2em}

\def\ppm{\times 10^{-6}}

\begin{document}
\begin{titlepage}
\pubblock

\vfill
\def\thefootnote{\fnsymbol{footnote}}
\Title{Status of the BNL Muon \g2 Experiment\support}
\vfill
\Author{R.~Prigl (2), H.N.~Brown (2), G.~Bunce (2),
R.M.~Carey (1), P.~Cushman (8),
G.T.~Danby (2), P.T.~Debevec (7), H.~Deng (12), W.~Deninger (7),
S.K.~Dhawan (12), V.P.~Druzhinin (9), L.~Duong (8), W.~Earle (1),
E.~Efstathiadis (1), F.J.M.~Farley (12), G.V.~Fedotovich (9),
S.~Giron (8), F.~Gray (7),
M.~Grosse Perdekamp (12), A.~Grossmann (5), U.~Haeberlen (6), M.~Hare (1),
E.S.~Hazen (1), D.W.~Hertzog (7),
V.W.~Hughes (12), M.~Iwasaki (11),
K.~Jungmann (5), D.~Kawall (12), M.~Kawamura (11), B.I.~Khazin (9),
J.~Kindem (8), F.~Krienen (1), I.~Kronkvist (8),
R.~Larsen (2), Y.Y.~Lee (2), W.~Liu (12), I.~Logashenko (9), R.~McNabb (8),
W.~Meng (2), J.-L.~Mi (2), J.P.~Miller (1),
W.M.~Morse (2), P.~Neumayer (5), D.~Nikas (2), C.J.G.~Onderwater (7), Y.~Orlov (3),
C.~Ozben (2), J.~Paley (1), C.~Pai (2), C.~Polly (7), J.~Pretz (12), G.~zu Putlitz (5),
S.I.~Redin (12), O.~Rind (1), B.L.~Roberts (1),
N.~Ryskulov (9), S.~Sedykh (7),
Y.K.~Semertzidis (2), Yu.M.~Shatunov (9), 
E.~Sichtermann (12), E.~Solodov (9), M.~Sossong (7),
A.~Steinmetz (12),
L.R.~Sulak (1), M.~Tanaka (2), C.~Timmermans (8),
A.~Trofimov (1), D.~Urner (7), D.~Warburton (2), D.~Winn (4),
A.~Yamamoto (10), D.~Zimmerman (8).}
\Address{(1) Department of Physics, Boston University, Boston, MA 02215,
USA,
(2) Brookhaven National Laboratory, Upton, NY 11973, USA,
(3) Newman Laboratory, Cornell University, Ithaca NY 14853, USA,
(4) Fairfield University, Fairfield, CT 06430, USA,
(5) Physikalisches Institut der Universit\"at Heidelberg, 69120
Heidelberg, Germany,
(6) MPI f\"ur Med.~Forschung, 69120 Heidelberg, Germany,
(7) Department of Physics, University of Illinois, Urbana, IL 61820, USA,
(8) Department of Physics, University of Minnesota, Minneapolis, MN 55455, USA,
(9) Budker Institute of Nuclear Physics, Novosibirsk, Russia,
(10) KEK, Japan,
(11) Tokyo Institute of Technology, Tokyo, Japan,
(12) Department of Physics, Yale University, New Haven, CT 06511, USA.
}
\vfill
\begin{Abstract}
The muon \g2 experiment at Brookhaven has been taking data
since 1997. Analyses of the data taken in 1997 and 1998, which
include about 2\% of the data taken so far, have improved the
experimental accuracy in the muon anomalous magnetic moment
to $a_{\mu (expt)} = 1 165 921(5) \times 10^{-9} (4\,ppm)$.
The value agrees with standard theory. Analysis of the 1999
data, about 25\% of the existing data set, is nearing completion
and analysis of the 2000 data has started. The experiment is
preparing for another major data taking run, this time storing
negative instead of positive muon beams.
\end{Abstract}
\vfill
\begin{Presented}
5th International Symposium on Radiative Corrections \\ 
(RADCOR--2000) \\[4pt]
Carmel CA, USA, 11--15 September, 2000
\end{Presented}
\vfill
\end{titlepage}
\def\thefootnote{\arabic{footnote}}
\setcounter{footnote}{0}
%


\section{Introduction}

The Brookhaven g-2 experiment E821 is designed to provide a precision test
of the standard model prediction for \amu which is dominated by the QED
radiative corrections but has sizeable contributions from hadronic loops,
$a_{\mu}^{Had} = 6739(67) \times 10^{-11}$ or $57.8(7)\, ppm$ in \amu,
as well as electroweak loops, $a_{\mu}^{EW} = 1.30(4)\, ppm$.
The theory of \amu is discussed
in detail in another contribution to these proceedings\hskip0.2em\cite{czarne}.
The goal of E821 is to reduce the error in \amu to $0.35\, ppm$, a fraction of
the electroweak contribution.

The experiment measures the muon anomaly \amu directly rather than the g-factor.
Therefore we commonly express contributions and errors in units of \amu. Any
numbers given in ppm here have to be multiplied by $1.165...\times 10^{-9}$ for
comparison with the absolute numbers found in\hskip0.2em\cite{czarne} and most
other theoretical papers on the muon g-factor.

\section{The Brookhaven Experiment}

The principle of the measurement
is similar to the third CERN experiment\hskip0.2em\cite{huai}.
Polarized muons are stored in a uniform dipole
magnetic field with electrostatic quadrupoles
providing weak
vertical focussing.  The muon spin precesses relative to
the momentum vector with the frequency
\be
\vec \omega_a = - {e \over m_\mu }\left[ a_{\mu} \vec B -
\left( a_{\mu}- {1 \over \gamma_\mu^2 - 1}\right)
\vec \beta \times \vec E \right],
\ee
where $\vec \beta  = \vec{v}/c$, $\gamma = 1/\sqrt{1-v^2/c^2}$,
and assuming that $\frac{E}{c} \ll B$ and $\vec{\beta}\cdot\vec{B} \approx 0$. 
The dependence of $\omega_a$ on the electric field $\vec{E}$
can be eliminated by storing
muons with the ``magic'' $\gamma_\mu$=29.3, corresponding to a muon momentum
$p_{\mu}$ = 3.094 GeV/$c$. In this ideal case,\linebreak[3]
$a_{\mu}-1/(\gamma_\mu^2-1)=0$, and the
focussing electric field does not affect the spin precession frequency.
$a_\mu$ is then
extracted from $\omega_a \approx 2\pi \times \mathrm{230\, kHz}$ through 
\be
a_\mu=\frac{ \omega_a/\omega_{\mathrm p} }{ \mu_\mu/\mu_{\mathrm p}
- \omega_a/\omega_{\mathrm p} }
\ee
where $\omega_{\mathrm p}\approx 2\pi \times \mathrm{62\, MHz}$ is the free
proton precession frequency in the same
magnetic field seen by the muons. The ratio of muon to proton magnetic 
moments is $\mu_\mu/\mu_{\mathrm p} = 3.183 345 39(10)$
\hskip0.2em\cite{PDG1,liu}.

The source of the stored muons is the Alternating Gradient Synchrotron
(AGS) proton beam, which delivers 6-12
bunches with a total of
40-60 $\times$~10$^{12}$ protons at 24 GeV/$c$ 
onto a nickel production target every 
2.7 s.
The individual bunches have a total width of about 100 ns and
are spaced apart by 33 ms.

From each bunch about $4\times 10^7$ pions at $\approx$ 3.1 GeV/$c$ 
are transported from the target\linebreak[3] along 
a 116 m beam line. About $\mathrm{50}\,$\%
of the pions decay along the transport line and
a momentum slit followed by
a bending magnet near the downstream end selects either pions or forward
decay muons from a slightly higher momentum pion beam for
injection into the storage ring.
After passing through
a hole in the back of the storage ring magnet yoke and a
field free region supplied by a
superconducting inflector magnet\hskip0.2em\cite{meng}, the pion or muon
beam enters the toroidal storage region which has a radius of 7.112 m
and a 9 cm diameter cross section. The storage ring magnet is described in
detail in\hskip0.2em\cite{danby}.

One of the major improvements over the last CERN experiment is the use
of direct muon injection.
In pion injection mode a small fraction of muons from pion decay,
$\pi^+ \rightarrow \mu^+ \nu_\mu$, are launched onto a stable
orbit and are stored.
In muon injection mode, a total kick of $\approx 11\, 
\mathrm{mrad}$
on the first one or two turns is needed to store the muons born in the pion
decay channel. The muon kicker is a pulsed magnet consisting of three sections
of pairs of current sheets, each 1.7 m long. The peak current through
the plates
during a 400 ns wide pulse
is about 4100 A, providing a vertical field of 0.016 T superimposed on the
1.45 T field of the storage ring magnet.

\begin{figure}
\center
\small
\vspace{-2.3cm}
\hspace*{-1cm}
\mbox
{ \epsfig{file=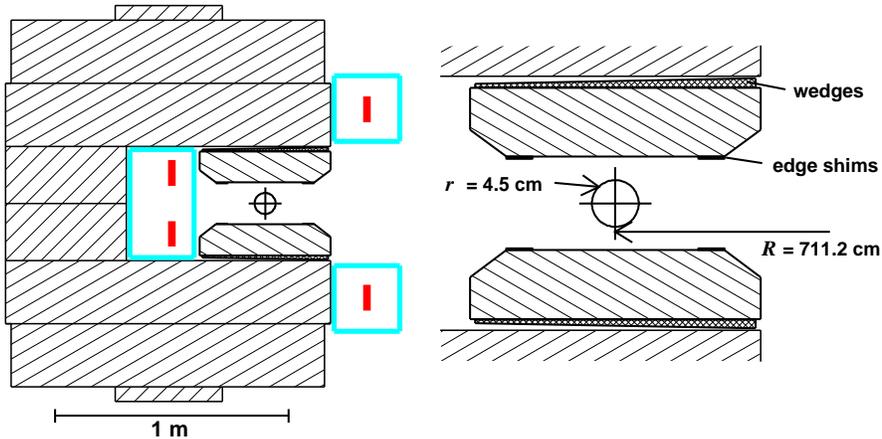,height=6in}}
\vspace{-7.1cm}

 \caption{\small Storage ring magnet cross section
and detailed view of the magnet gap region.
}
\label{fig:777}
\end{figure}

\begin{figure*}[p]
\center
\small
\vspace{-2.0cm}
\mbox
{ \epsfig{file=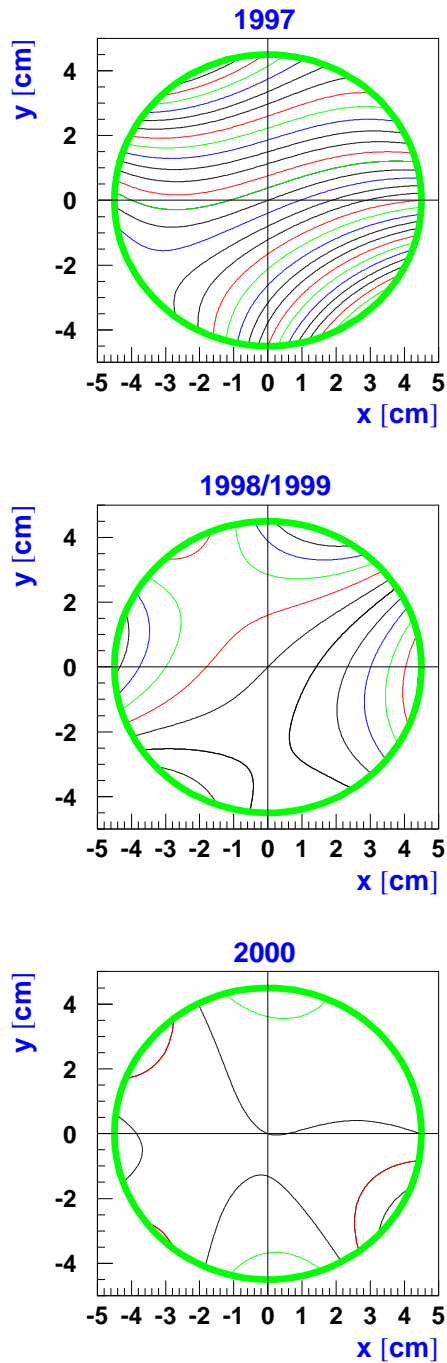,height=8in}}
\caption{\small Typical contour plot of the magnetic field integrated
over azimuth during the 1997 run (top), 1998 (1999) run (middle), and the
2000 run (bottom). {\it x} denotes the radial and {\it y} the
vertical direction. Each contour line represents a fractional change
of $1\ppm$. No efforts were made to improve the field quality between
the 1998 and 1999 run and the field quality was the same for these runs.
The most significant improvement for 2000 came from the installation of
a new inflector magnet.
{\em ~~~~~~~~~~~~~~~~ \qquad
~~~~~~~~~~~~~~ 
\qquad}}
\label{fig:778}
\end{figure*}

The continuous superferric `C'-shaped storage ring magnet,
 Fig.~\ref{fig:777},
is excited by
superconducting coils which carry a current of $5177$ A.
The yoke consists of twelve 30 degree sections bolted together
at the four corners, with azimuthal gaps of less than 1 mm. The pole
pieces are 10 degrees long and aligned with the yoke sectors.
The azimuthal gap between adjacent pole pieces of about $75\,
\mathrm{\mu m}$ is filled with insulating Kapton foils to avoid irregular eddy
current effects. The vertical gap between pole and yoke decouples
the yoke and pole pieces, which are fabricated from high quality
steel, and allows the insertion of iron wedges to compensate for
the C-magnet quadrupole.
The 10 cm wide wedges
are radially adjustable. This allows us to locally change the total
air gap and thus the dipole field component for better field homogeneity in
azimuth. The four edge shims, 5 cm wide and about 3 mm
high, are the main tool for
reducing field variations over the beam cross section. 
Continuous current sheets glued to the pole pieces are used to
further reduce the fractional inhomogeneity in the integral field.
The gradual improvement of the field integral across the aperture
is shown in Fig.~\ref{fig:778}.

\begin{figure*}[t]
\center
\small
\hspace*{-0.5cm}
\vspace{-2.5cm}
\mbox
{ \epsfig{file=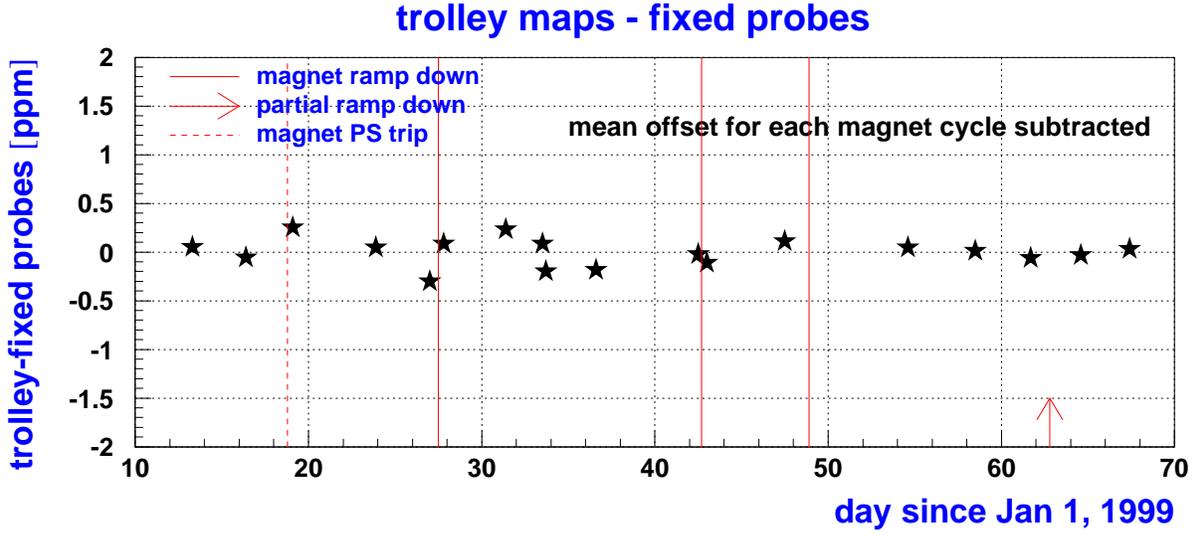,height=3.4in}}
\vspace{1.3cm}
\caption{\small Difference between the field average
measured by the trolley and the field average predicted
by the monitoring probes. It is assumed that the true difference
is constant during each magnet-on period and fluctuations are caused
by the limited accuracy of the individual measurements.
{\em ~~~~~~~~~~~~~~~~ \qquad
~~~~~~~~~~~~~~ 
\qquad}}
\label{fig:779}
\end{figure*}

During data taking an array of nuclear magnetic resonance (NMR) probes
embedded in the top and bottom plates of the twelve vacuum chambers
is used to monitor and stabilize the magnetic field\hskip0.2em\cite{ralf}.
About one third of the 375 probes installed are typically used in
the field analysis. 
The field inside the storage region is mapped twice
a week using a hermetically sealed trolley operating in vacuum and
containing a matrix of 17 NMR probes. The probes
in the trolley are calibrated in place against a standard
probe\hskip0.2em\cite{fei}. The first trolley run in a magnet cycle
establishes the offset between the average field measured by the
monitoring probes and the average field seen by the muon beam.
Subsequent trolley runs can then be used to determine the accuracy
of the continuous field ``tracking'' with the monitoring probes.
The quality of the field tracking is shown in Fig.~\ref{fig:779}
for the 1999 run.

The decay positrons from $\mu^+ \rightarrow \mathrm{e^+ \nu_{e} \bar
\nu_{\mu}}$, which
constitute our signal, range in energy from 0 GeV to 3.1 GeV,
and are 
detected with 24 Pb-scintillating fiber calorimeters 
placed symmetrically around the inside of the storage 
ring. Because of the parity violating nature of the weak decay,
the high-energy positrons are preferentially emitted
along the muon spin direction. The muon spin precession is reflected
in the decay positron spectrum, $N(t)$, where we expect
\be
 N(t)=N_0(E)e^{-t/{\tau_\mu}}\left[1-A(E)\cos\left(\omega_a t+\phi\right)\right].
\label{5pfit}
\ee
The normalization constant $N_0$ depends on the energy threshold E as does
the asymmetry parameter A. For E=2.0 GeV, $A$ is $\approx 0.4$.

The arrival times of the positrons are recorded in multi-hit
time-to-digital converters (TDCs), and the calorimeter pulses are also
sampled by a 400 MHz waveform digitizer (WFD).
A laser and light emitting diode (LED) system are
used to monitor potential time and gain shifts.
Several detector stations are outfitted with
a finely segmented hodoscope array of 20$\times$32 small scintillating
elements connected to a multianode
phototube, which provide position sensitive information on the muon 
decay positron. Additional event information
is derived from stations
equipped with five scintillator paddles oriented horizontally. 
Finally, a set of wire chambers in one section of the ring
provides information on the stored muon phase
space by measuring the flight path of the decay electrons and
tracing them back to their origin. The distribution of the muon beam
across the aperture has to be convoluted
with the magnetic field to determine 
the average field seen by the muons.
The mean radial distribution of the stored muons can also be obtained from
the fast rotation signal from the initial time structure of the injected
beam.

\section{Present Status of Results}

In the first data taking run of the experiment in 1997 pion injection
was used because the muon kicker was still under construction. This mode
suffers from a low efficiency - only about 20 muons get stored per $10^6$ pions
injected into the storage ring - as well as from a significant flash caused
by hadronic interactions of pions that do not decay and hit the inflector channel
wall at the end of the first turn. The result from this run, $a_{\mu} = 1\,165\,925(15)
\times 10^{-9}$, is discussed in\hskip0.2em\cite{1997res}. In 1998 the
muon kicker was commissioned and first data were taken in muon
injection mode. The run was cut short by a hardware failure in
the beamline. The analysis of the 1998 data, most of which taken during
the last week of running, resulted in a value of $a_{\mu} = 11\,659\,191(59)
\times 10^{-10}\, (\pm 5\,ppm) $\hskip0.2em\cite{1998res}.
Combining all experimental data including
the old CERN measurement, $a_{\mu}^{CERN} = 11\,659\,230(84)
\times 10^{-10}\, (\pm 7.2\,ppm)$\hskip0.2em\cite{huai},
 yields $a_{\mu}^{expt} = 11\,659\,205(46)
\times 10^{-9}\, (\pm 4\,ppm) $ for the world average. This value
agrees with the theoretical value $a_{\mu}^{SM} = 11\,659\,160(7)
\times 10^{-10}\, (\pm 0.6\,ppm)$ found in\hskip0.2em\cite{czarne}.

\section{Status of the Analysis of the 1999 and 2000 Data}

\begin{figure*}[t]
\center
\small
\hspace*{-0.5cm}
\vspace{-8.5cm}
\mbox
{ \epsfig{file=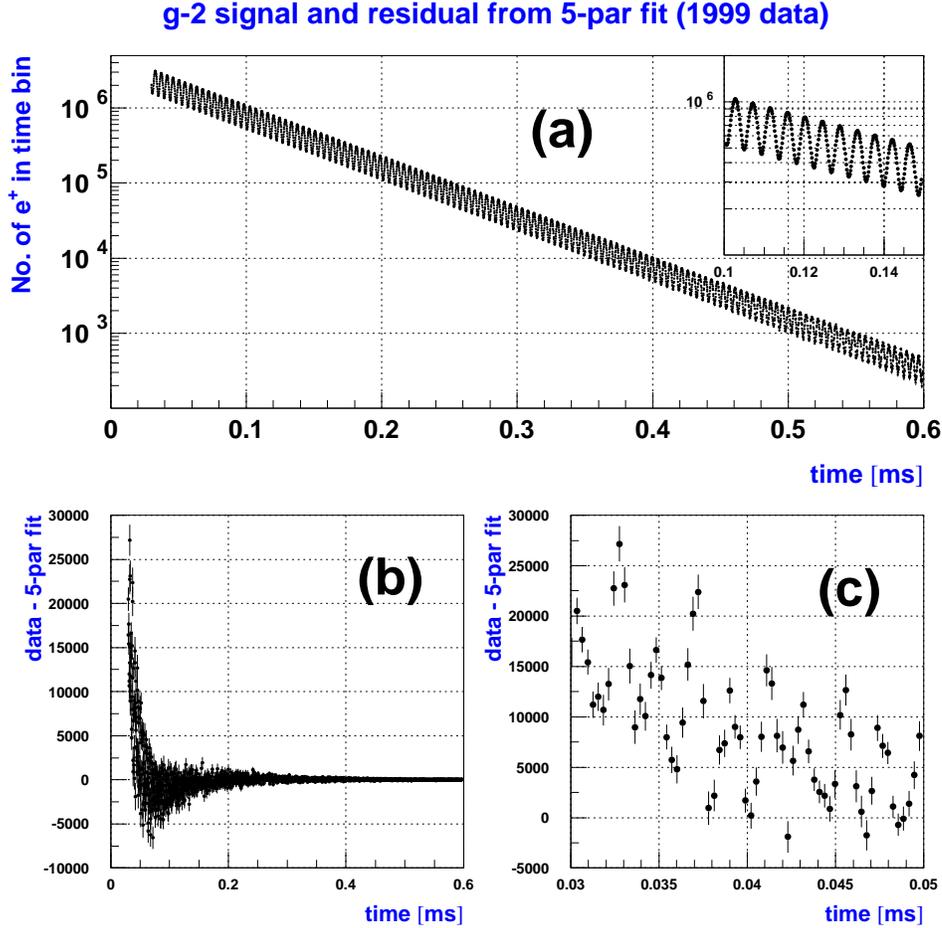,height=5.6in}}
\vspace{7.3cm}
\caption{\small Positron spectrum from a large fraction of
the 1999 data set (a), with an enlarged view of the time
between 100 and 150 $\mu sec$ after beam injection in the top right
corner. The residuals from fitting the function in eq. (\ref{5pfit})
to the data show clear structure (b), particularly at early times (c).
{\em ~~~~~~~~~~~~~~~~ \qquad
~~~~~~~~~~~~~~ 
\qquad}}
\label{fig:782}
\end{figure*}

The experiment currently analyses the data collected in 1999 which
is about 15 times larger than the 1998 data set. In contrast to the
old data, which could quite well be fitted to the basic 5-parameter
function in equation (\ref{5pfit}), the significant increase in statistical power combined
with a higher data taking rate revealed a number of subtle effects that
slowed down the analysis. Two of the most prominent effects can be illustrated
by ignorantly fitting the data to the function in eq. (\ref{5pfit}), and then
looking at the residuals, i.e. the difference between the measured decay
spectrum and the 5-parameter fit  as shown in Fig.~\ref{fig:782}.
Clearly the fitting function does not describe the data well.
The residual in Fig.~\ref{fig:782}b,c has two prominent features.
One is the excess of data at early times, and the other an oscillation
of the counts with respect to the fitting function, also decaying
with time. The former feature is caused by the high counting rate at
early times. Our 400 MHz waveform digitizer does not allow us to
distinguish between two decay positrons that hit the same detector
within about 3.5 nsec. Such pairs of pulses pile up to one larger
pulse. If each of the single pulses exceed the
energy threshold E, we lose a count. On the other hand, two pulses
below threshold can combine to form a larger pulse that exceeds the
threshold. Since our energy cuts are relatively high to benefit from
the large asymmetry at higher energies, the pulses gained outnumber
the pulses lost, leading to excessive counts at early times. An additional
complication is that the g-2 phase of the pulses lost is not the same
as the phase of the pulses gained because higher energy decay positrons have
a larger bend radius in the ring magnet field and on average a longer
flight path and flight time before hitting one of the calorimeters.
Choosing an energy
threshold where the pulses gained and pulses lost cancel does not
eliminate the pile-up problem.
As the muons decay, the count rate and therefore the fraction of
pile-up pulses goes down. The simple 5-parameter fit to the data
tries to accomodate the higher counts at early times by
increasing the normalization constant, leading to an overestimate of
counts at later times as seen in Fig.~\ref{fig:782}b.

An elegant way to
extract the number of pile-up pulses from the data is to artificially increase the
deadtime of the pulse finder by a factor of two, see how many more pulses
are found above threshold, and then subtract these counts from the
original decay positron spectrum. This sample of pulses can also
be used to study the phase of the pile-up pulses. An alternative is to
parametrize the pile-up fraction and include it in the fitting function.
Both methods are used and compared in the analysis.

\begin{figure*}[t]
\center
\small
\vspace{-0.5cm}
\mbox
{ \epsfig{file=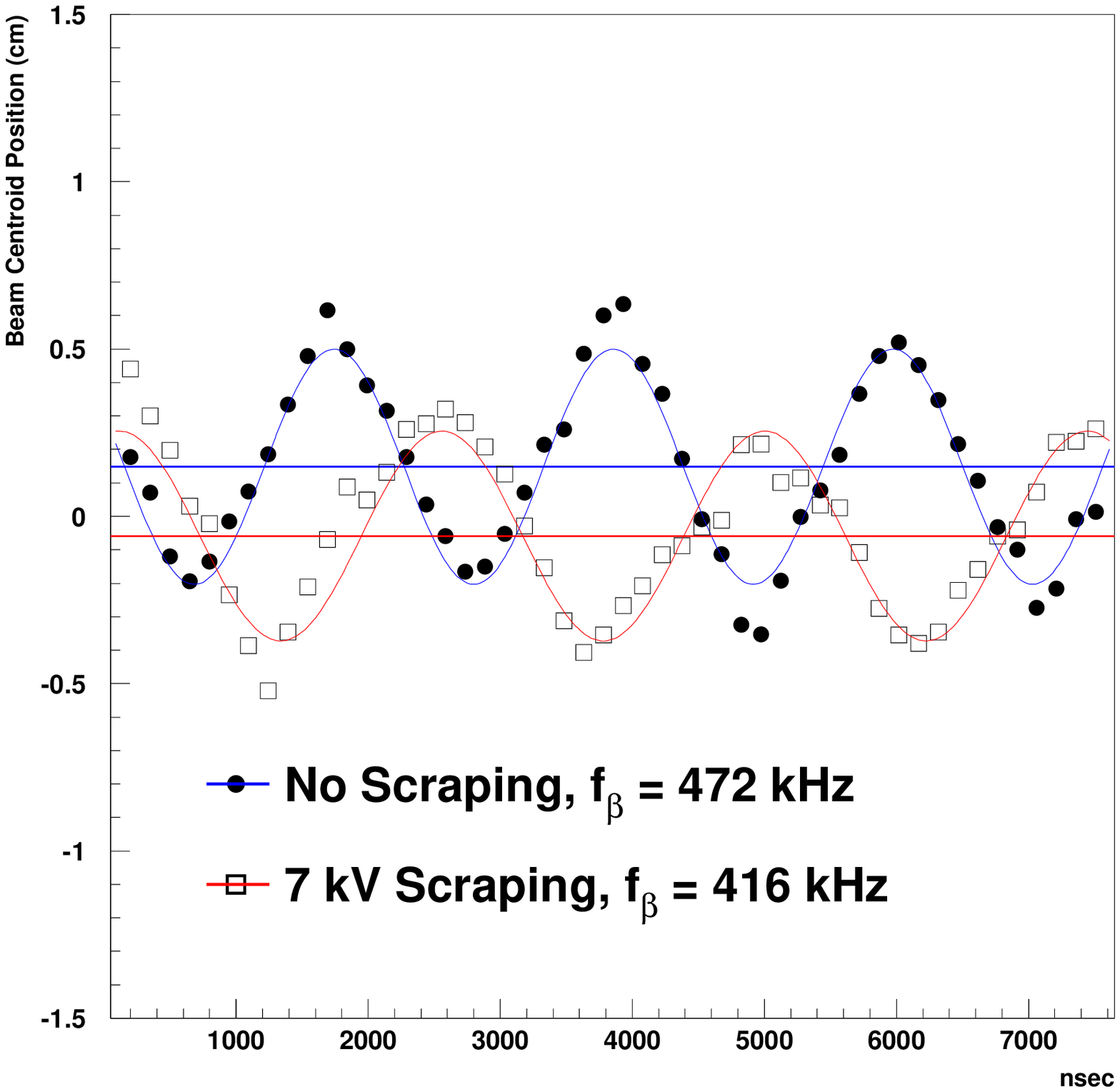,height=4in}}
\caption{\small Beam centroid as measured by a set of
fibers plunged into the muon beam at two different
settings of the electrostatic quadrupoles. In normal
running mode, the muon beam gets scraped for the first
15-20 $\mu$sec after injection by an imbalance in the
quadrupole plate voltages which lowers the center
of the electrostatic quadrupole field and therefore the
beam centroid. This also changes the betatron tune.
{\em ~~~~~~~~~~~~~~~~ \qquad
~~~~~~~~~~~~~~ 
\qquad}}
\label{fig:780} 
\end{figure*}

The oscillations in the residuals are caused by a coherent
motion of the stored muon beam with a frequency
that depends on the strength of
the focusing electrostatic quadrupole field, the betatron tune.
Due to a mismatch between the narrow inflector channel aperture,
an 18\hskip0.2emmm horizontal times 56\hskip0.2emmm vertical
rectangle, and the 90\hskip0.2emmm diameter circular ring magnet
aperture, we do not fill the available phase space in the storage
ring in muon injection mode. This leads to a modulation of the
horizontal beam width at the betatron oscillation frequency
from the momentum spread, and a modulation of the horizontal
and vertical beam width at twice this frequency from the angular
spread of the incoming beam. In addition, our muon kicker runs
reliably only a few percent below its design value. The resulting
incomplete kick, combined with the fact that we do not fill the
storage ring phase space, causes the beam centroid to oscillate
about the central orbit at the betatron frequency. This coherent
beam motion is known as \underline{c}oherent
\underline{b}etatron \underline{o}scillation or CBO.
The dynamic behaviour of the stored muon beam was
measured with a set of fiber harps
that plunge in and out of the beam. Fig.~\ref{fig:780} shows
the motion of the beam centroid as an example.

The CBO is visible in the positron spectrum because the
acceptance of our detectors is slightly dependent on the
radius of the muon at the time of decay.
The frequency we see is not the CBO frequency
itself but its beating with the frequency of revolution, since we
detect the decays at fixed points along the storage ring magnet
circumference.

Other effects that need to be studied in detail include the
loss of muons before they decay, the gain stability of the calorimeters
and timing stability of the readout electronics, background events caused
by the loss of protons that we inevitably store together with the muons,
and background events called ``flashlets''. The flashlets are caused
by a small fraction of the halo of the primary beam still circulating
in the accelerator being scattered into the extraction line and
transported to the production target. The resulting low intensity
secondary beam does not get kicked and is therefore not stored
in our ring magnet, but creates a small flash of background signals.
The most troublesome secondary beam component is positrons. Starting
with the momentum of the stored muon beam, the positrons spiral
inwards due to energy loss in the inflector channel windows and give
a signal that cannot be distinguished from true decay positron
counts near the upper edge of the decay spectrum. We have developed
a sensitive analysis method to detect these flashlets utilizing the discrete
time structure associated with the $2.7\, \mu$sec revolution time
of the beam in the accelerator. The flashlet contamination is
very sensitive to the performance of the accelerator and we
disregard data taken while there were elevated levels of
flashlet contamination. During the 2000 run we installed a sweeper
magnet in the secondary beam line that turns on microseconds
after beam injection into the storage ring and sweeps any
late arrivals out of the beamline.

In the course of the data analysis all these effects
have to be studied in detail, with different analysis teams
developing their own tools to correct for them, parametrize
them for inclusion into the fitting function, or estimate
their impact on the fit parameters with analytical and Monte
Carlo methods in cases where the unwanted or missing counts
do not have unique characteristics but mix into other fitting
parameters. This process is nearing completion for the 1999 data.

\section{Future}

The experiment is looking forward to another run from February through
April 2001. Since the 2000 run we have reversed the polarity of the beam line
and ring magnet to store negative muon beam and measure $a_{\mu^-}$.
We expect a decay positron rate similar to
the 2000 run and hope to at least double our data set. Analysis of the
2000 data, 3-4 times larger than the 1999 data set,
has started and initial checks e.g. for flashlet contamination
suggest that most of the data is of good quality.

\end{document}